\documentclass[aip,twocolumn]{revtex4}

\usepackage{amsmath}
\usepackage{amssymb}
\usepackage{stmaryrd}

\usepackage{graphicx}

\usepackage{epsfig}

\def\ov#1{\overline{#1}}

\def\vb#1{\mbox{\boldmath$#1$}}

\def\wh#1{\widehat{#1}}

\def\btimes{\,\vb{\times}\,}

\newcommand{\bc}{\begin{center}}
\newcommand{\ec}{\end{center}}
\newcommand{\bt}{\begin{tabbing}}
\newcommand{\et}{\end{tabbing}}
\newcommand{\be}{\begin{eqnarray*}}
\newcommand{\ee}{\end{eqnarray*}}
\newcommand{\bs}{\begin{slide}}
\newcommand{\es}{\end{slide}}

\begin{document}

\title{Notes on the exact solutions and singularities of an X-point collapse \\ in Hall magnetohydrodynamics}

\author{Alain J.~Brizard}
\affiliation{Department of Physics, Saint Michael's College, Colchester, VT 05439, USA}

\begin{abstract}
A recent paper by A.~Z.~Janda [J.~Math.~Phys.~{\bf 59}, 061509 (2018)] presented a partial solution of the equations of dissipationless Hall magnetohydrodynamics (MHD) in terms of the Weierstrass elliptic function. Unfortunately, an error crept up in the analysis, where the kinetic and potential energies of the initial Hall MHD state were misidentified. The present comment presents the correct energy analysis and offers a complete solution of dissipationless Hall magnetohydrodynamics  in terms of the Jacobi elliptic functions.
\end{abstract}

\maketitle

In a recent paper \cite{Janda_2018}, Janda investigated the solutions to a 2D model based on the equations of dissipationless Hall magnetohydrodynamics (MHD). The model (also previously studied by Litvinenko \cite{Litvinenko_2007} and Litvinenko \& McMahon \cite{Litvinenko_2015}) involves the coupling between the incompressible fluid velocity ${\bf V} = \wh{\sf z}\btimes\nabla\phi + V_{z}\,\wh{\sf z}$, which is expressed in terms of the electrostatic potential $\phi(x,y,t) = \gamma(t)\,xy$ and the parallel fluid velocity $V_{z}(x,y,t) = \beta_{1}(t)\,x^{2} + \beta_{2}(t)\,y^{2}$, and the magnetic field ${\bf B} = \nabla\psi\btimes\wh{\sf z} + B_{z}\,\wh{\sf z}$, which is expressed in terms of the parallel vector potential $\psi(x,y,t) = \alpha_{1}(t)\,x^{2} - \alpha_{2}(t)\,y^{2}$ and the parallel magnetic field $B_{z}(x,y,t) = b(t)\;x\,y$. Here, the Hall MHD fields $(\phi,V_{z},\psi,B_{z})$ are dimensionless and $(x,y)$ have been normalized to a characteristic length scale $L$. The time-dependent coefficients $(\alpha_{1},\alpha_{2},\beta_{1},\beta_{2},b)$ satisfy the coupled ordinary differential equations \cite{Janda_2018}
\begin{eqnarray}
\dot{\alpha}_{1} - 2\gamma\,\alpha_{1} & = & 2b\,({\sf d}_{i}\alpha_{1} - {\sf d}_{e}^{2}\beta_{1}), \label{eq:alpha_1} \\
\dot{\alpha}_{2} + 2\gamma\,\alpha_{2} & = & -\;2b\,({\sf d}_{i}\alpha_{2} + {\sf d}_{e}^{2}\beta_{2}), \label{eq:alpha_2} \\
\dot{\beta}_{1} - 2\gamma\,\beta_{1} & = & -\;2\,b\,\alpha_{1}, \label{eq:beta_1} \\
\dot{\beta}_{2} + 2\gamma\,\beta_{2} & = & -\;2\,b\,\alpha_{2}, \label{eq:beta_2} \\
\dot{b} & = & -\,4\;(\alpha_{1}\,\beta_{2} \;+\; \alpha_{2}\,\beta_{1}), \label{eq:b}
\end{eqnarray}
with $\gamma$ appearing as an unconstrained coefficient since the vorticity $\nabla^{2}\phi \equiv 0$ in the model under consideration. In Eqs.~\eqref{eq:alpha_1}-\eqref{eq:alpha_2}, the plasma parameters
${\sf d}_{i,e} = \sqrt{m_{i,e}c^{2}/(4\pi ne^{2}L^{2})}$ denotes the normalized ion/electron skin depth, where $m_{i,e}$ denotes the ion/electron mass and we assume a quasineutral plasma with equal ion/electron densities: $n_{i} = n = n_{e}$. 

Equations \eqref{eq:alpha_1}-\eqref{eq:b} have the following conservation laws 
\be
\alpha_{1}\,\alpha_{2} - \alpha_{1}^{0}\,\alpha_{2}^{0}  & = &  \frac{1}{4}\,{\sf d}_{e}^{2}\left(b^{2} \;-\; b_{0}^{2}\right) \\
\beta_{1}\,\beta_{2} - \beta_{1}^{0}\,\beta_{2}^{0} & = &  \frac{1}{4}\left(b^{2} \;-\; b_{0}^{2}\right)  \\
\left(\alpha_{1} + {\sf d}_{i}\,\beta_{1}\right) \left( \alpha_{2} - {\sf d}_{i}\,\beta_{2}\right)  & = & \left(\alpha_{1}^{0} + {\sf d}_{i}\,\beta_{1}^{0}\right) \left( \alpha_{2}^{0} - {\sf d}_{i}\,
\beta_{2}^{0}\right) \\
 &  &+\;  \frac{1}{4}\,{\sf d}_{e}^{2}\left(b^{2} \;-\; b_{0}^{2}\right) 
\ee
where $\alpha_{1,2}(0) = \alpha_{1,2}^{0}$, $\beta_{1,2}(0) = \beta_{1,2}^{0}$, and $b(0) = b_{0}$ denote initial conditions. If we take the time derivative of Eq.~\eqref{eq:b}, and substitute Eqs.~\eqref{eq:alpha_1}-\eqref{eq:beta_2}, we obtain a second-order differential equation for $b(t)$, with initial conditions $b(0) = b_{0}$ and $\dot{b}(0) = \dot{b}_{0}$:
\begin{equation} 
\ddot{b} \;-\; 2\,\left(4\,{\sf d}_{e}^{2} + {\sf d}_{i}^{2}\right)\,b^{3} \;+\; 2\,{\sf c}\; b \;=\; 0,
\label{eq:b_ddot}
\end{equation}
where the initial-data coefficient ${\sf c}$ is defined as
\be 
{\sf c} & = & 4\,{\sf d}_{i}\,\left(\alpha_{1}^{0}\beta_{2}^{0} \;-\; \alpha_{2}^{0}\beta_{1}^{0}\right) \;+\; \left(4\,{\sf d}_{e}^{2} + {\sf d}_{i}^{2}\right)\;b_{0}^{2} \\
 &  &-\; 8 \left(\alpha_{1}^{0}\alpha_{2}^{0} \;+\; {\sf d}_{e}^{2}\;\beta_{1}^{0}\beta_{2}^{0}\right).
\ee
Next, we multiply Eq.~\eqref{eq:b_ddot} by $2\,\dot{b}$ and integrate to obtain
\[ \dot{b}^{2} - \dot{b}_{0}^{2} = \left(4{\sf d}_{e}^{2} + {\sf d}_{i}^{2}\right) \left( b^{4} - b_{0}^{4} \right) - 2\,{\sf c} \left( b^{2} - b_{0}^{2}\right), \]
where $\dot{b}_{0}^{2} = 16(\alpha_{1}^{0}\,\beta_{2}^{0} + \alpha_{2}^{0}\,\beta_{1}^{0})^{2}$ is obtained from Eq.~\eqref{eq:b}.

We simplify this equation by introducing the normalization: $q(t) \equiv \sqrt{4\,{\sf d}_{e}^{2} + {\sf d}_{i}^{2}}\;b(t)$, so that we obtain the energy conservation law
\begin{equation} 
\frac{1}{2}\,\dot{q}^{2} \;+\; U(q) \;=\; E \;=\; \frac{1}{2}\,\dot{q}_{0}^{2} \;+\; U(q_{0}),
\label{eq:Energy_q}
\end{equation}
where $E$ denotes the total energy of the system and the potential energy 
\begin{equation} 
U(q) \;=\; {\sf c}\,q^{2} \;-\; \frac{1}{2}\,q^{4},
\label{eq:Uq}
\end{equation}
has two maxima ${\sf c}^{2}/2$ at $q = \pm\,\sqrt{{\sf c}}$ and a local minimum at $q = 0$. In Eq.~(28) of Janda's work \cite{Janda_2018}, the relation $\dot{q}^{2} = q^{4} - 2{\sf c}q^{2} + ({\sf c}^{2} - {\sf c}_{0}^{2})$ is introduced with the definition given by Janda's Eq.~(29): ${\sf c}_{0}^{(J)} = \sqrt{(q_{0}^{2} - {\sf c})^{2} \;+\; \dot{q}_{0}^{2}}$. However, if we use the initial condition $\dot{q}_{0}^{2} = (4{\sf d}_{e}^{2} + {\sf d}_{i}^{2})\,\dot{b}_{0}^{2}$ obtained from Janda's Eq.~(21), we arrive at the erroneous relation $\dot{q}^{2} = (q^{4} - q_{0}^{4}) - 2{\sf c}(q^{2} - q_{0}^{2}) - \dot{q}_{0}^{2}$, which violates the initial condition $q_{0} = q(0)$ and $\dot{q}_{0} = \dot{q}(0)$. The sign error is easily corrected with the new definition
\begin{equation}
{\sf c}_{0} \equiv \sqrt{(q_{0}^{2} - {\sf c})^{2} \;-\; \dot{q}_{0}^{2}},
\label{eq:c0}
\end{equation}
from which we recover the energy conservation law \eqref{eq:Energy_q}.

 \begin{figure}
\epsfysize=1.8in
\epsfbox{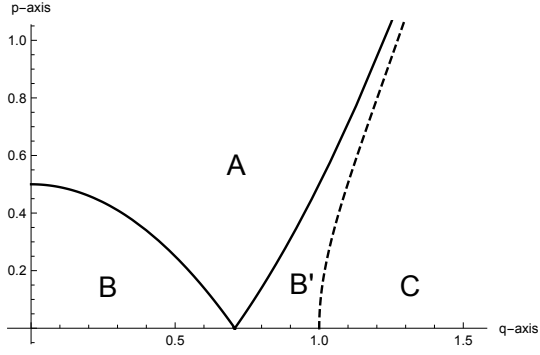}
\caption{Phase-space portrait $(q, p= \dot{q})$ for ${\sf c} = 1/2$. The solid curve corresponds to the separatrix solution at $\epsilon = 1$, while the dashed curve corresponds to $\epsilon = 0$. Solutions in region A $(\epsilon \geq 1)$ correspond to singular unbounded orbits. Region B $(0 \leq \epsilon \leq 1)$ is divided into two subregions: periodic orbits below the separatrix (B) and singular unbounded orbits to the right of the separatrix (B$^{\prime}$). Solutions in region C correspond to singular unbounded orbits with $\epsilon \leq 0$.}
\label{fig:portrait}
\end{figure}

We now proceed with a turning-point analysis of Eq.~\eqref{eq:Energy_q}. First, we introduce the energy normalization: $E \equiv \frac{1}{2}\,{\sf c}^{2}\epsilon$, where $\epsilon$ is a dimensionless real parameter, and the parameter ${\sf c}_{0}$, which is now correctly defined in Eq.~\eqref{eq:c0}, is related to ${\sf c}$ as ${\sf c}_{0} = {\sf c}\sqrt{1 - \epsilon}$. We note that Janda's work \cite{Janda_2018} only investigates the range 
$0 \leq \epsilon \leq 1$, where ${\sf c}_{0} \leq {\sf c}$ (which follows immediately from its correct definition). By using the energy conservation law \eqref{eq:Energy_q}, with Eq.~\eqref{eq:c0}, we obtain the general Jacobi elliptic equation \cite{NIST_Jacobi} 
\begin{equation} 
\dot{q}^{2} \;=\; {\sf c}^{2}\epsilon \left( 1 \;-\; \frac{q^{2}}{{\sf c} + {\sf c}_{0}}\right) \left( 1 \;-\; \frac{q^{2}}{{\sf c} - {\sf c}_{0}}\right),
\label{eq:Jacobi}
\end{equation}
which is naturally solved in terms of the Jacobi elliptic functions (as suggested by Litvinenko \cite{Litvinenko_2007} and Litvinenko \& McMahon \cite{Litvinenko_2015}). A unified treatment of the turning-point analysis is carried out with $\epsilon \equiv \sin^{2}\varphi$, where
\begin{equation}
\varphi = \left\{ \begin{array}{llr}
(A) & \pi/2 + i\,\nu & (\nu \geq 0) \\
(B) & \mu & (0 \leq \mu \leq \pi/2) \\
(C) & i\,\nu & (\nu \geq 0)
\end{array} \right.
\label{eq:ABC}
\end{equation}
allow us to obtain an explicit solution for each region (A: $\epsilon \geq 1$; B: $0 \leq \epsilon \leq 1$; C: $\epsilon \leq 0$). In region A, there are no turning points since $\epsilon = \cosh^{2}\nu \;\geq\; 1$, i.e., the energy level is above the two maxima and the solution represents an unbounded orbit. Hence, it is convenient to choose the initial condition $q_{0} = 0$ and $\dot{q}_{0} = \pm\,{\sf c}\,\cosh\nu$ for an unbounded orbit initially moving either left or right. In region B, where $0 \leq \epsilon = \sin^{2}\mu \leq 1$, there are four real turning points, where we choose the initial condition $\dot{q}_{0} = 0$ and either $q_{0} = \pm\,\sqrt{2{\sf c}}
\cos(\mu/2)$ for an unbounded orbit, or $q_{0} = \pm\,\sqrt{2{\sf c}}\sin(\mu/2)$ for a periodic bounded orbit. Lastly, in region C, where $\epsilon = -\,\sinh^{2}\nu$, there are two real turning points with initial conditions $\dot{q}_{0} = 0$ and $q_{0} = \pm\,\sqrt{2{\sf c}}\cosh(\nu/2)$ for an unbounded orbit.

\begin{figure}
\epsfysize=1.8in
\epsfbox{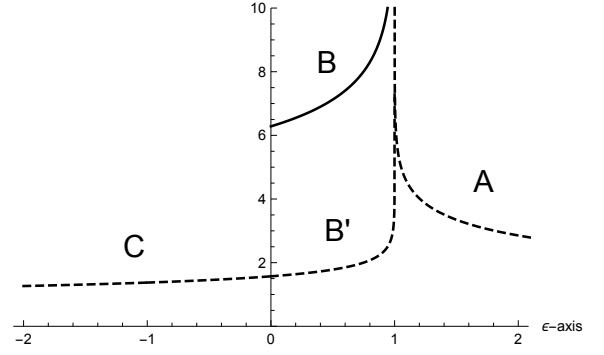}
\caption{Plots of the period $T({\sf c},\epsilon)$ (solid curve B in the range $0 \leq \epsilon \leq 1$), the singularity finite-time $T_{\infty}({\sf c},\epsilon) = \frac{1}{4}\,T({\sf c},\epsilon)$ (dashed curves B$^{\prime}$ and C in the range $-2 \leq \epsilon \leq 1$), and the singularity finite time $T_{\infty}({\sf c},\epsilon) = {\rm Re}[2 {\sf K}(k)/\Omega(\nu)]$ (dashed curve A in the range $1 \leq \epsilon \leq 2$) for ${\sf c} = 1/2$. We note that all times become infinite as $\epsilon \rightarrow 1$ (corresponding to the separatrix solution).}
\label{fig:finite_time}
\end{figure}

The solutions (31) presented by Janda \cite{Janda_2018} for regions B and C are expressed in terms of the Weierstrass elliptic function $\wp(t; g_{2}, g_{3})$:
\begin{equation}
q_{\pm}(t) \;=\; q_{0}^{\pm} \left(1\pm \frac{{\sf c}_{0}}{\wp(t) - {\sf e}_{1} \mp {\sf c}_{0}/2}\right),
\label{eq:wp_sol}
\end{equation} 
where $q_{0}^{\pm} = \sqrt{{\sf c} \pm {\sf c}_{0}}$ is a turning point (${\sf c}_{0}$ is real in these regions), and  the upper sign in Eq.~\eqref{eq:wp_sol} corresponds to the unbounded orbits in regions B and C, while the lower sign corresponds to the periodic bounded orbits in region B.  In addition, the Weierstrass elliptic function $\wp(t; g_{2}, g_{3})$, which has a singularity at $t = 0$, satisfies the differential equation \cite{NIST_W} $\dot{\wp}^{2} = 4\,(\wp - {\sf e}_{1})(\wp - {\sf e}_{2})(\wp - {\sf e}_{3})$, where the cubic root ${\sf e}_{1} = \frac{1}{3}\,{\sf c}$ is energy-independent, while the other two cubic roots ${\sf e}_{2,3}(\epsilon) = -\,
\frac{1}{6}\,{\sf c} \pm \frac{1}{2}\,{\sf c}\sqrt{\epsilon}$ are energy-dependent, so that ${\sf e}_{1} \geq {\sf e}_{2} \geq {\sf e}_{3}$ (for $0 \leq \epsilon \leq 1$) and ${\sf e}_{1} + {\sf e}_{2} + {\sf e}_{3} = 0$ (for all values 
of $\epsilon$). The periodic solution (for $0 \leq \epsilon \leq 1$) has the orbital period
\begin{equation}
T({\sf c},\epsilon) \;=\; 2\,\omega_{1}({\sf c},\epsilon) \;\equiv\; 2\,{\sf K}(k)/\sqrt{{\sf e}_{1} - {\sf e}_{3}},
\label{eq:T}
\end{equation}
where $\omega_{1}$ denotes the real-valued half period of $\wp(t)$, with $\wp(\omega_{1}) = {\sf e}_{1}$, ${\sf e}_{1} - {\sf e}_{3} = U^{\prime\prime}(0)\,(1 + \sqrt{\epsilon})/4$, with $U^{\prime\prime}(0) = 2\,{\sf c}$, and ${\sf K}(k)$ denotes the complete elliptic integral of the first kind \cite{NIST_EK}, with $k^{2}(\epsilon) = ({\sf e}_{2} - {\sf e}_{3})/({\sf e}_{1} - {\sf e}_{3}) = 2\,\sqrt{\epsilon}/(1 + \sqrt{\epsilon})$. At the quarter period $T/4 = \omega_{1}/2$, we use formula 23.7.1 of Ref.~\cite{NIST_W}
\[ \wp(\omega_{1}/2) = {\sf e}_{1} + \sqrt{({\sf e}_{1} - {\sf e}_{2})({\sf e}_{1} - {\sf e}_{3})} = {\sf c}/3 + {\sf c}_{0}/2, \]
to show that the periodic solution $q_{-}(\omega_{1}/2) = 0$, while $q_{+}(\omega_{1}/2) = \infty$, i.e., the unbounded solution has a singularity at a finite time $T_{\infty}({\sf c},\epsilon) = \omega_{1}({\sf c},\epsilon)/2$. The period $T({\sf c},\epsilon)$, defined in Eq.~\eqref{eq:T} and the singularity finite-time $T_{\infty}({\sf c},\epsilon) = \frac{1}{4}T({\sf c},\epsilon)$ are shown in Fig.~\ref{fig:finite_time} as a solid curve (B) and a dashed curve (B$^{\prime}$ and C), respectively, for ${\sf c} = 1/2$. The unbounded solution in region A $(\epsilon \geq 1)$ also has a finite-time singularity (shown in Fig.~\ref{fig:finite_time} as a dashed curve A).

We now obtain Jacobi elliptic solutions of Eq.~\eqref{eq:Jacobi} for each region defined in Eq.~\eqref{eq:ABC}, which will be useful in obtaining explicit solutions for $\alpha_{1,2}$ and $\beta_{1,2}$. First, for region A, we find ${\sf c} \pm {\sf c}_{0} = {\sf c}\,\exp(\pm i\Phi)\cosh\nu$, where $\Phi = \arctan(\sinh\nu)$. Using the initial condition: $q_{0} = 0$ and $\dot{q}_{0} = {\sf c}\,\cosh\nu$, we obtain the singular unbounded solution
\begin{equation}
q_{A}(t) \;=\; \sqrt{\dot{q}_{0}}\;e^{i\Phi/2}\;{\rm sn}(\Omega t, k),
\label{eq:orbit_A}
\end{equation}
where $\Omega \equiv \sqrt{\dot{q}_{0}}\exp(-i\Phi/2)$ and $k \equiv \exp(i\Phi)$. We note that the solution \eqref{eq:orbit_A} is real for all real times, and reaches infinity at the finite time \footnote{The poles of 
${\rm sn}(z,k)$ are located at $(2n+1)\,i{\sf K}(k') + 2m\,{\sf K}(k)$, for $(m,n) \in \mathbb{Z}$. In our case, however, ${\rm Re}[i\,{\sf K}(k')/\Omega(\nu)] = 0$ so the pole at $i{\sf K}(k') + 2\,{\sf K}(k)$ is considered.} $T_{\infty}({\sf c},\epsilon) = {\rm Re}[2 {\sf K}(k)/\Omega(\nu)]$, while its velocity $\dot{q} = \dot{q}_{0}\,{\rm cn}(\Omega t, k){\rm dn}(\Omega t, k)$ does not vanish for real times and reaches a minimum at ${\sf c}\,\sqrt{\epsilon - 1} = {\sf c}\,\sinh\nu$. In addition, we note that the Weierstrass solution in region A is expressed as
\begin{equation}
q_{A}(t) \;=\; e^{i\Phi/2}\sqrt{\wp\left(e^{i\Phi/2}\,t + \omega_{3}\right) \;-\; {\sf e}_{3}},
\label{eq:qA_wp}
\end{equation}
with $\dot{q}_{A}(t) = \exp(i\Phi)\sqrt{(\wp - {\sf e}_{1})(\wp - {\sf e}_{2})}$. Here, the cubic roots $({\sf e}_{1}, {\sf e}_{2},{\sf e}_{3})$ are 
\begin{equation}
\left. \begin{array}{rcl}
{\sf e}_{1} & = & -\dot{q}_{0}(1 - 2\, e^{-2i\Phi})/3 \\
{\sf e}_{2} & = & \dot{q}_{0}(2 - e^{-2i\Phi})/3 \\
{\sf e}_{3} & = & -\dot{q}_{0}(1 + e^{-2i\Phi})/3
\end{array} \right\},
\end{equation}
so that $\dot{q}_{A}(0) = \exp(i\Phi)\sqrt{({\sf e}_{3} - {\sf e}_{1})({\sf e}_{3} - {\sf e}_{2})} = \dot{q}_{0}$. We note that the separatrix solution between $\pm\,\sqrt{{\sf c}}$ with 
$(q_{0},\dot{q}_{0}) = (0,{\sf c})$ is obtained from Eq.~\eqref{eq:orbit_A} in the limit $\nu \rightarrow 0$: $q_{\rm sp}(t) = \sqrt{{\sf c}}\;\tanh(\sqrt{{\sf c}}\;t)$, where we used ${\rm sn}(u,1) = \tanh(u)$. Here, we see that the separatrix solution has an infinite period since $q_{\rm sp}(t) \rightarrow \sqrt{{\sf c}}$ only as $t \rightarrow \infty$ (see the limit $\epsilon \rightarrow 1$ in Fig.~\ref{fig:finite_time}).

\begin{figure}
\epsfysize=1.8in
\epsfbox{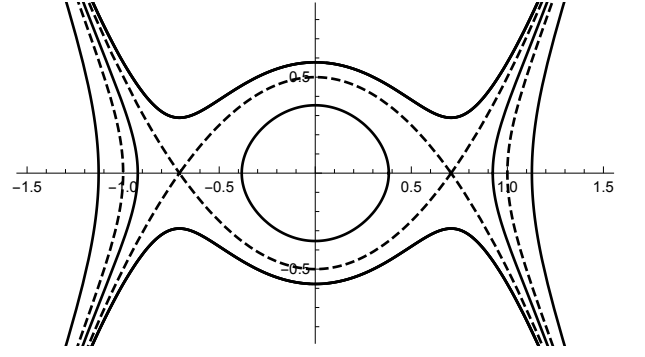}
\caption{Sample phase-space portrait $(q, p= \dot{q})$ for ${\sf c} = 1/2$, showing solutions \eqref{eq:orbit_A}, \eqref{eq:orbit_BI}, \eqref{eq:orbit_BII}, and \eqref{eq:orbit_C}.}
\label{fig:portrait_sample}
\end{figure}

Second, we consider the periodic bounded-orbit solutions in region B, where ${\sf c} \pm {\sf c}_{0} = {\sf c}\,(1 \pm \cos\mu)$. Using the initial condition: $\dot{q}_{0} = 0$ and $q_{0}^{-} = \sqrt{2\,{\sf c}}\,
\sin(\mu/2)$, we obtain $q_{B}^{-}(t) = q_{0}\,{\rm sn}(\Omega t + {\sf K}(k), k)$, where $\Omega = \sqrt{2\,{\sf c}}\,\cos(\mu/2)$ and $k = \tan(\mu/2) \leq 1$. Using Table 22.4.3 of Ref.~\cite{NIST_Jacobi}, we obtain the periodic solution
\begin{equation}
q_{B}^{-}(t) \;=\; q_{0}^{-}\;{\rm cn}(\Omega t, k)/{\rm dn}(\Omega t, k).
\label{eq:orbit_BI}
\end{equation}
Since ${\rm cn}(\Omega t, k)/{\rm dn}(\Omega t, k)$ is periodic and has no singularities along the real-time axis, the orbital period is
\begin{equation}
T({\sf c},\mu) \;=\; \frac{4{\sf K}(k)}{\Omega} \;=\; \frac{4\,{\sf K}(\tan(\mu/2))}{\sqrt{U^{\prime\prime}(0)}\;\cos(\mu/2)},
\label{eq:T_mu}
\end{equation}
which yields the simple-harmonic-oscillator limit $T({\sf c},0) = 2\pi/\sqrt{U^{\prime\prime}(0)}$ at $\epsilon = 0$ (at the origin of the phase-space portrait in Fig.~\ref{fig:portrait_sample}). We note that the periods \eqref{eq:T} and \eqref{eq:T_mu} are identical as a result of the descending Landen transformation (see formula 19.8.12 of Ref.~\cite{NIST_EK}): ${\sf K}(k) = (1 + k_{1})\,{\sf K}(k_{1})$, where $k^{2} = 2\sin\mu/(1 + \sin\mu)$ and $k_{1} = \tan(\mu/2)$, with $1 + k_{1} = \sqrt{1 + \sin\mu}/\cos(\mu/2)$ and $1 + \sin\mu = [\cos(\mu/2) + \sin(\mu/2)]^{2}$.

Region B also has an unbounded-orbit solution $q_{B}^{+}(t) = q_{0}^{+}\;{\rm cn}(\Omega t, k)/{\rm dn}(\Omega t, k)$, with the initial condition: $\dot{q}_{0} = 0$ and $q_{0}^{+} = \sqrt{2\,{\sf c}}\,\cos(\mu/2)$, with 
$\Omega = \sqrt{2\,{\sf c}}\,\sin(\mu/2)$ and $k = \cot(\mu/2) \geq 1$. Since $k > 1$, we use the formulas 22.17.3-4 of Ref.~\cite{NIST_Jacobi}, and we find
\begin{equation}
q_{B}^{+}(t) \;=\; q_{0}^{+}\;{\rm dn}(\Omega k\;t, k^{-1})/{\rm cn}(\Omega k\;t, k^{-1}).
\label{eq:orbit_BII}
\end{equation}
Since this solution becomes infinite when ${\rm cn}$ vanishes (i.e., when $t = {\sf K}/\Omega$), we find the singularity finite-time
\begin{equation}
T_{\infty}({\sf c},\mu) \;=\; \frac{{\sf K}(\tan(\mu/2))}{\sqrt{2\,{\sf c}}\;\cos(\mu/2)},
\end{equation}
which is one quarter of the orbital period \eqref{eq:T_mu}. 

Lastly, region C has an unbounded-orbit solution $q_{C}(t) = q_{0}^{+}\;{\rm cn}(i\,\Omega t, k)$, with initial condition: $\dot{q}_{0} = 0$ and $q_{0}^{+} = \sqrt{2\,{\sf c}}\,\cosh(\nu/2)$, where $\Omega = \sqrt{2\,{\sf c}\,\cosh\nu}$ and $k = \cosh(\nu/2)/\sqrt{\cosh\nu} \leq 1$. Using Table 22.6.1 of Ref.~\cite{NIST_Jacobi}, we find
\begin{equation}
q_{C}(t) \;=\; q_{0}^{+}/{\rm cn}(\Omega t, k^{\prime}),
\label{eq:orbit_C}
\end{equation}
where $k^{\prime 2} = 1 - k^{2} =  \sinh^{2}(\nu/2)/\cosh\nu$, which also has a singularity finite-time
\[ T_{\infty}({\sf c},\nu) \;=\; \frac{{\sf K}(\sinh(\nu/2)/\sqrt{\cosh\nu})}{\sqrt{2\,{\sf c}\,\cosh\nu}}. \]
With the solutions \eqref{eq:orbit_A}, \eqref{eq:orbit_BI}, \eqref{eq:orbit_BII}, and \eqref{eq:orbit_C} for $q(t) \equiv \sqrt{{\sf d}_{i}^{2} + 4{\sf d}_{e}^{2}}\;b(t)$ now completely known (some sample solutions are shown in Fig.~\ref{fig:portrait_sample}), we can explore the solutions for the remaining time-dependent coefficients $\alpha_{1,2}(t)$ and $\beta_{1,2}(t)$ governed by Eqs.~\eqref{eq:alpha_1}-\eqref{eq:beta_2}.

We now follow Janda's analysis \cite{Janda_2018} and we discuss the solution of Eqs.~\eqref{eq:alpha_1}-\eqref{eq:beta_2} in the limit ${\sf d}_{e} = 0$:
\begin{eqnarray}
\dot{\ov{\alpha}}_{1}  & = & 2\,q\;\ov{\alpha}_{1}, \label{eq:alpha_10} \\
\dot{\ov{\alpha}}_{2}  & = & - 2\,q\;\ov{\alpha}_{2}, \label{eq:alpha_20} \\
{\sf d}_{i}\dot{\ov{\beta}}_{1} & = & - 2\,q\;\ov{\alpha}_{1}, \label{eq:beta_10} \\
{\sf d}_{i}\dot{\ov{\beta}}_{2} & = &  - 2\,q\;\ov{\alpha}_{2}, \label{eq:beta_20}
\end{eqnarray}
with the definitions
\begin{eqnarray}
\left(\begin{array}{c}
\alpha_{1}(t) \\
\beta_{1}(t)
\end{array}\right) & = & \left(\begin{array}{c}
\ov{\alpha}_{1}(t) \\
\ov{\beta}_{1}(t)
\end{array}\right) e^{2\Gamma(t)}, \\
\left(\begin{array}{c}
\alpha_{2}(t) \\
\beta_{2}(t)
\end{array}\right) & = & \left(\begin{array}{c}
\ov{\alpha}_{2}(t) \\
\ov{\beta}_{2}(t)
\end{array}\right) e^{-\,2\Gamma(t)},
\end{eqnarray}
where $\Gamma(t) \equiv \int_{0}^{t} \gamma(t') dt'$. The solution of Eqs.~\eqref{eq:alpha_10}-\eqref{eq:alpha_20} is expressed as
\begin{equation}
\left(\begin{array}{c}
\ov{\alpha}_{1}(t) \\
\ov{\alpha}_{2}(t)
\end{array}\right) \;=\; \left(\begin{array}{c}
\alpha_{1}^{0} \\
\alpha_{2}^{0}
\end{array}\right) \exp\left( \pm\frac{}{} 2 Q(t)\right),
\label{eq:alpha_12_sol}
\end{equation}
where $Q(t) \equiv \int_{0}^{t} q(t') dt'$ and $(\ov{\alpha}_{1}^{0},\ov{\alpha}_{2}^{0}) = (\alpha_{1}^{0},\alpha_{2}^{0})$. Next, inserting Eq.~\eqref{eq:alpha_12_sol} into Eqs.~\eqref{eq:beta_10}-\eqref{eq:beta_20}, we find
\begin{eqnarray}
\left(\begin{array}{c}
\dot{\ov{\beta}}_{1}(t) \\
\dot{\ov{\beta}}_{2}(t)
\end{array}\right)  & = & \mp \left(\begin{array}{c}
\alpha_{1}^{0}/{\sf d}_{i} \\
\alpha_{2}^{0}/{\sf d}_{i}
\end{array}\right) \left( \pm\,2\,q\frac{}{}e^{\pm 2Q}\right) \nonumber \\
 & = & \mp \left(\begin{array}{c}
\alpha_{1}^{0}/{\sf d}_{i} \\
\alpha_{2}^{0}/{\sf d}_{i}
\end{array}\right) \frac{d\left(e^{\pm 2Q}\right)}{dt},
\label{eq:beta_12}
\end{eqnarray}
whose solutions are
\begin{eqnarray}
\ov{\beta}_{1}(t) & = & \beta_{1}^{0} + \alpha_{1}^{0}/{\sf d}_{i} \left(1 \;-\frac{}{} e^{2Q}\right), \label{eq:beta_1_sol} \\
\ov{\beta}_{2}(t) & = & \beta_{2}^{0} - \alpha_{2}^{0}/{\sf d}_{i} \left(1 \;-\frac{}{} e^{-2Q}\right). \label{eq:beta_2_sol}
\end{eqnarray}
Hence, the general solutions for $\alpha_{1,2}$ and $\beta_{1,2}$ are 
\begin{eqnarray}
\alpha_{1,2}(t) & = & \alpha_{1,2}^{0}\;\exp\left(\pm\,2\;\Gamma(t) \pm\frac{}{} 2\,Q(t)\right), \label{eq:alpha_12_sol} \\
\beta_{1,2}(t) & = & \left( \beta_{1,2}^{0} \pm \frac{\alpha_{1,2}^{0}}{{\sf d}_{i}}\right) e^{\pm2\;\Gamma(t)} \;\mp\; \frac{\alpha_{1,2}(t)}{{\sf d}_{i}},
\label{eq:beta_12_sol}
\end{eqnarray}
where the upper sign corresponds to $(\alpha_{1},\beta_{1})$ and the lower sign corresponds to $(\alpha_{2},\beta_{2})$.

Lastly, we can now easily evaluate the integral $Q(t)$ for the solutions \eqref{eq:orbit_A}, \eqref{eq:orbit_BI}, \eqref{eq:orbit_BII}, and \eqref{eq:orbit_C} in terms of standard formulas found in Sec.~2.7 of Ref.~\cite{Lawden}. First, for the singular unbounded orbit \eqref{eq:orbit_A}, we find
\begin{eqnarray}
Q_{A}(t) & = & k \int_{0}^{\Omega t} {\rm sn}(u,k) du \nonumber \\
 & = & \ln\left( \frac{{\rm dn}(\Omega t, k) - k\,{\rm cn}(\Omega t, k)}{1 - k}\right),
\end{eqnarray}
where $k = \exp(i \Phi)$ and $\Omega = \sqrt{\dot{q}_{0}}\,\exp(-i\Phi/2)$, so that
\begin{equation}
e^{2Q_{A}(t)} \;=\; \left( \frac{{\rm dn}(\Omega t, k) - k\,{\rm cn}(\Omega t, k)}{1 - k}\right)^{2}.
\label{eq:eQA}
\end{equation} 
Next, for the periodic bounded orbit \eqref{eq:orbit_BI}, we find
\begin{equation}
Q_{B}^{-}(t) = k \int_{0}^{\Omega t} \frac{{\rm cn}(u,k)}{{\rm dn}(u,k)} du = \ln\left( \frac{1 + k\,{\rm sn}(\Omega t, k)}{{\rm dn}(\Omega t, k)}\right),
\end{equation}
where $k = \tan(\mu/2)$ and $\Omega = \sqrt{2{\sf c}}\,\cos(\mu/2)$, so that
\begin{equation}
e^{2Q_{B}^{-}(t)} \;=\; \left( \frac{1 + k\,{\rm sn}(\Omega t, k)}{{\rm dn}(\Omega t, k)}\right)^{2}.
\label{eq:eQBI}
\end{equation} 
For the singular unbounded orbit \eqref{eq:orbit_BII}, we find
\begin{equation}
Q_{B}^{+}(t) = \int_{0}^{k\Omega t} \frac{{\rm dn}(u,k)}{{\rm cn}(u,k)}\;du = \ln\left( \frac{1 + {\rm sn}(k\Omega t, k^{-1})}{{\rm cn}(k\Omega t, k^{-1})}\right),
\end{equation}
where $k^{-1} = \tan(\mu/2)$ and $k\Omega = \sqrt{2{\sf c}}\,\cos(\mu/2)$, so that
\begin{equation}
e^{2Q_{B}^{+}(t)} \;=\; \left( \frac{1 + {\rm sn}(k\Omega t, k^{-1})}{{\rm cn}(k\Omega t, k^{-1})}\right)^{2}.
\label{eq:eQBII}
\end{equation} 
Lastly, for the singular unbounded orbit \eqref{eq:orbit_C}, we find
\begin{eqnarray}
Q_{C}(t) & = & k \int_{0}^{\Omega t} \frac{du}{{\rm cn}(u,k')} \nonumber \\
 & = & \ln\left( \frac{{\rm dn}(\Omega t, k') + k\,{\rm sn}(\Omega t, k')}{{\rm cn}(\Omega t, k')}\right),
\end{eqnarray}
where $k^{\prime} = \sinh(\nu/2)/\sqrt{\cosh\nu}$, $k = \cosh(\nu/2)/\sqrt{\cosh\nu}$, and $\Omega = \sqrt{2{\sf c}\,\cosh\nu}$, so that
\begin{equation}
e^{2Q_{C}(t)} \;=\; \left( \frac{{\rm dn}(\Omega t, k') + k\,{\rm sn}(\Omega t, k')}{{\rm cn}(\Omega t, k')}\right)^{2}.
\label{eq:eQC}
\end{equation} 
We note that, while the exponential factor \eqref{eq:eQBI} reaches a maximum $[1+\tan(\mu/2)]/[1-\tan(\mu/2)]$ at the quarter period, the exponential factors \eqref{eq:eQA}, \eqref{eq:eQBII}, and \eqref{eq:eQC} become infinite at their respective singularity finite times $T_{\infty}$. Hence, according to Eqs.~\eqref{eq:alpha_12_sol}-\eqref{eq:beta_12_sol}, $(\alpha_{1},\beta_{1})$ become infinite at the singularity finite-times $T_{\infty}$, while $\alpha_{2} \rightarrow 0$ and $\beta_{2} \rightarrow (\beta_{2}^{0} - \alpha_{2}^{0}/{\sf d}_{i}) \exp[-2\;\Gamma(T_{\infty})]$. While the solutions for $b(t)$ were obtained to all orders in ${\sf d}_{e}$, corrections to the lowest-order coefficients \eqref{eq:alpha_12_sol}-\eqref{eq:beta_12_sol} can be obtained by perturbation method.

\end{document}